\documentclass[twocolumn,twoside,nofootinbib,prd,aps,tightenlines,10pt]{revtex4-1}

\usepackage{amssymb}
 
\usepackage{graphicx}
\usepackage{latexsym}
\usepackage{mathrsfs}
\usepackage{calrsfs}
\usepackage{amsfonts}
\usepackage{color}

{\count255=\time\divide\count255 by 60 \xdef\hourmin{\number\count255}
  \multiply\count255 by-60\advance\count255 by\time
  \xdef\hourmin{\hourmin:\ifnum\count255<10 0\fi\the\count255}}

\def\vev#1{ \left\langle #1 \right \rangle }

\begin{document}

\title{Large Scale Anomalies in the Microwave Background: Causation and Correlation}

\author{Grigor Aslanyan}\email{g.aslanyan@auckland.ac.nz}
\author{Richard Easther}

\affiliation{Department of Physics, University of Auckland,
  Private Bag 92019, Auckland, New Zealand }

\begin{abstract}
Most treatments of large scale  anomalies in the microwave sky  are   {\em a posteriori\/}, with  unquantified look-elsewhere effects. We contrast these with physical models of specific inhomogeneities in the early Universe which can generate these apparent anomalies. Physical models  predict correlations {\em between\/}  candidate anomalies and the corresponding signals in polarization and large scale structure, reducing the impact of cosmic variance. We compute the apparent spatial curvature associated with large-scale inhomogeneities and show that it is typically small, allowing for a self-consistent analysis. As an illustrative example we show that a single large plane wave inhomogeneity  can contribute to low-$l$ mode alignment and odd-even asymmetry in the power spectra and the best-fit model accounts for a significant part of the claimed odd-even asymmetry.  We argue that this approach can be generalized to provide a more quantitative assessment of potential large scale anomalies in the Universe. 
\end{abstract}

\maketitle

 {\em Planck} \cite{Ade:2013ktc} and {\em WMAP} \cite{Bennett:2012fp}  yield independent, consistent maps of the cosmic microwave background (CMB).    At large scales the main origin of uncertainty is the finite size of the visible Universe, rather than instrumental sensitivity or integrated observing time.  Many CMB anomalies have been described \cite{Bennett:2010jb,Ade:2013nlj,Land:2005ad,Tegmark:2003ve,deOliveiraCosta:2003pu,Rakic:2007ve,Aslanyan:2013lsa,ArmendarizPicon:2010da,Aslanyan:2013zs,Land:2005jq,Kim:2010gf,Gruppuso:2010nd}, including low-$l$ mode alignments or ``axis of evil'', cold spot, hemispherical asymmetry, and parity asymmetry.  These analyses are {\em a posteriori\/}  and  subject to large, unquantified ``look-elsewhere'' effects,  and correlations  between candidate anomalies are unknown.    Further,   $p$ values and similar  tests are often used to assess the significance of anomalies \cite{Ade:2013nlj}, even though we cannot  observe an ensemble of microwave skies.      
 
 By contrast,  a physical mechanism that generates large-scale anisotropy will also predict the   correlations between candidate anomalies, including those for which a search of our sky yields a null result,  along with the  associated polarization pattern.  As an illustrative example, we analyze a model with a single amplified Fourier mode, showing this can generate  low-$l$ mode alignments and parity asymmetry in the CMB \cite{Ade:2013nlj}. We  compute the expected curvature $\Omega_K$ generated by a large perturbation \cite{Bull:2013fga}, finding that perturbations large enough to generate anomalies in the CMB do not also generate an apparent spatial curvature.

We work within the broad inflationary paradigm, but similar arguments could be made in other scenarios.  Assume for simplicity that the perturbations have the Bunch-Davies form \cite{Baumann:2009ds}, the Hubble parameter $H\approx \mathrm{const}$, and the spectral index  $n_s =1$.  Single-field  inflation \cite{Baumann:2009ds} with gauge-invariant curvature perturbations $\zeta$ on uniform density hypersurfaces has the Mukhanov variable  $v\equiv z\zeta$, 
\begin{equation}
z^2\equiv a^2\frac{\dot{\phi}^2}{H^2}=2a^2\epsilon\,,
\end{equation}
where $a$ is the scale factor, $\phi$ is the inflaton, and $\epsilon$ is the slow-roll parameter.   We quantize $v$  via
\begin{equation}
\hat{v}_\mathbf{k}=v_\mathbf{k}(\tau)\hat{a}_{\mathbf{k}}+v^*_{-\mathbf{k}}(\tau)\hat{a}^\dagger_{-\mathbf{k}}
\end{equation}
where $\hat{a}^\dagger$ and $\hat{a}$ are the usual creation and annihilation operators, $\tau$ is the conformal time and 
\begin{equation}
v(\tau, \mathbf{x})=\int\frac{d^3k}{(2\pi)^3}v_{\mathbf{k}}(\tau)e^{i\mathbf{k}\cdot\mathbf{x}} \, .
\end{equation}
The mode functions $v_\mathbf{k}$  satisfy the  Mukhanov equation.

The Bunch-Davies vacuum state $|0\rangle$ obeys
\begin{equation}
\hat{a}_\mathbf{k}|0\rangle=0,\;\;\lim_{\tau\rightarrow-\infty}v_\mathbf{k}=\frac{e^{-ik\tau}}{\sqrt{2k}}
\end{equation}
implying
\begin{equation}
v_\mathbf{k}=\frac{e^{-ik\tau}}{\sqrt{2k}}\left(1-\frac{i}{k\tau}\right)\,.
\end{equation}
At superhorizon scales the one- and two-point functions of the gauge-invariant curvature perturbation $\zeta$ are
\begin{equation}
\vev{\zeta_\mathbf{k}} = 0\,, \quad \vev{\zeta_\mathbf{k}\zeta_{\mathbf{k}^\prime}}=(2\pi)^3\delta(\mathbf{k}+\mathbf{k}^\prime)\frac{H_*^2}{2k^3}\frac{H_*^2}{\dot{\phi}_*^2}\, ,
\end{equation}
where  $*$ denotes a value at horizon crossing. Now assume that some modes have particle content, for which the $v_\mathbf{k}$ will differ from their vacuum values. For simplicity, consider a coherent state $|\psi\rangle$, i.e.
\begin{equation}
\hat{a}_\mathbf{k}|\psi\rangle=\alpha_\mathbf{k}|\psi\rangle\,.
\end{equation}
If $\vev{\zeta_\mathbf{k}}=0$ at superhorizon scales  the coherent  Bunch-Davies states lead to identical two-point and three-point functions  \cite{Kundu:2011sg}. However,  for a general coherent  state
%
\begin{eqnarray}\label{zeta1point}
\vev{\zeta_\mathbf{k}}&=&-\frac{H^2}{\dot{\phi}^2\sqrt{2k^3}} \times \nonumber \\ &&\quad
\left(e^{-ik\tau}(k\tau-i)\alpha_\mathbf{k}+e^{ik\tau}(k\tau+i)\alpha^*_{-\mathbf{k}}\right)\,,  \\
\label{zeta2point}
\vev{\zeta_\mathbf{k}\zeta_{\mathbf{k}^\prime}}&=&4\pi^3\delta(\mathbf{k}+\mathbf{k}^\prime)\frac{H^4}{k^3 {\dot{\phi}^2}} (1+k^2\tau^2)+\vev{\zeta_\mathbf{k}}\vev{\zeta_{\mathbf{k}^\prime}}\!.
\end{eqnarray}
One may think of $\zeta$ as consisting of a quantum and a semiclassical piece \cite{Aslanyan:2013zs}, and we treat $\vev{\zeta_\mathbf{k}}$ as a semiclassical fluctuation where the $\alpha_\mathbf{k}$ are constrained by the  data.

This scenario is  a nonminimal model of inflation, in that the perturbations are not strictly Bunch-Davies. This is unsurprising, given that the purported anomalies  are defined by their relatively low likelihood of being generated within the simplest models for the very early Universe.  Given that this model is intended only as an illustrative example we do not provide a detailed mechanism for generating the additional perturbation. However, departures from Bunch-Davies will be common in models where the overall period of inflation is short and initial inhomegeneities are not fully erased.

Given  we are interested in long wavelength curvature perturbations, we first ask whether these  generate an effective spatial curvature  $\Omega_K$ within our Hubble volume if inflation starts at $\tau_s$ with the one-point function $\vev{\zeta(\mathbf{x})} \sim 1$ and $\Omega_K \sim 1$ when $\tau = \tau_s$ and we  assume that currently observable modes were well inside the horizon at $\tau_s$, or $k\tau_s\gg1$. The one-point functions (\ref{zeta1point})  are  approximately 
\begin{equation}
\vev{\zeta_\mathbf{k}}_s\simeq-\frac{H^2\tau_s}{\dot{\phi}^2\sqrt{2k}}\left(e^{-ik\tau_s}\alpha_\mathbf{k}+e^{ik\tau_s}\alpha^*_{-\mathbf{k}}\right)\,.
\end{equation}
When inflation  begins $\vev{\zeta_\mathbf{k}} \sim 1/k^3$, since $\zeta$ is dimensionless in real space, and $\zeta_\mathbf{k}$ has units of $1/k^3$, so
\begin{equation}\label{alpha_magnitude}
\alpha_\mathbf{k}\sim\frac{\dot{\phi}^2\sqrt{2k}}{H^2\tau_sk^3}\,.
\end{equation}

For superhorizon modes $k\tau\ll1$, and $\vev{\zeta_\mathbf{k}}$   has the limit
\begin{equation}
\vev{\zeta_\mathbf{k}}_\mathrm{super}\simeq\frac{H^2}{\dot{\phi}^2\sqrt{2k}k}i\left(\alpha_\mathbf{k}-\alpha^*_{-\mathbf{k}}\right)\,.
\end{equation}
Using the order-of-magnitude result, Eq.~(\ref{alpha_magnitude}),
\begin{equation}
\vev{\zeta_\mathbf{k}}_\mathrm{super}\sim\frac{1}{k^4\tau_s}=A_f\frac{1}{k\tau_f}\frac{1}{k^3}\,.
\end{equation}
A mode $k_f$ crosses the horizon at $\tau_f$. We define $A_f$, a dimensionless measure of the perturbation at a scale $k_f$
\begin{equation}
A_f=\frac{\tau_f}{\tau_s}\,,
\end{equation}
which vanishes in the limit that inflation has lasted for an arbitrarily long period.  

Assuming rapid thermalization  the Hubble parameter during inflation gives the density at the onset  of radiation domination. The scale factor evolves from  $a_s$ to $a_\mathrm{end}$ during inflation and again taking $\Omega_K\sim1$ and recalling that $\Omega_K\propto 1/a^2H^2$   \cite{Baumann:2009ds}, we find
\begin{equation}
\Omega_{K,\mathrm{end}}\sim\left(\frac{a_s}{a_\mathrm{end}}\right)^2=\left(\frac{A_f}{\tau_f}\right)^2\frac{1}{a_\mathrm{end}^2H_\mathrm{end}^2}\,,
\end{equation}
and, at  present, 
\begin{equation}\label{omegak_amplitude}
\Omega_{K,0}=\Omega_{K,\mathrm{end}}\frac{a_\mathrm{end}^2H_\mathrm{end}^2}{a_0^2H_0^2}\sim\left(\frac{A_fk_f}{a_0H_0}\right)^2
\end{equation}
where a subscript  $0$ denotes a present-day value. We thus find a simple relationship between the present value of $\Omega_K$, the initial fluctuation amplitude $A_f$, and the comoving scale $k_f$.  Furthermore, $\Omega_{K,0}$ is proportional to $A_f^2$ and does not depend on the scale of inflation. 

Recognizing $R_\mathrm{curv}=a^{-1}H^{-1}/\sqrt{|\Omega_K|}$, the curvature radius, in Eq. (\ref{omegak_amplitude})  we find
\begin{equation}
\frac{1}{k_f}\sim A_fR_\mathrm{curv}\,.
\end{equation}
Large values of $A_f$ are incompatible with CMB data \cite{Aslanyan:2013zs} implying $A_f\ll1$, so we are considering fluctuations with a wavelength much smaller than the curvature radius.    


As a specific, illustrative example now consider a fluctuation in a single mode with the form
\begin{equation}\label{my_model}
\vev{\zeta(\mathbf{x})}_{super}=A_f\cos(\mathbf{k_f}\cdot\mathbf{x}+\alpha)\,,
\end{equation}
with $A_f \sim1$ when inflation starts.  This scenario is discussed in  \cite{Aslanyan:2013zs}, which shows that $A_f \lesssim10^{-4}$ for scales  larger than $0.6\,L_0$ where $L_0=14.4\,\text{Gpc}$, the radius of the last scattering surface. Consequently, for this scenario $|\Omega_{K,0}| \lesssim 10^{-7}$. If $\Omega_K \sim 1$ at the onset of inflation we can inject fluctuations into the early Universe which  match CMB anomalies and ensure that $\Omega_K$ is compatible with current bounds, $|\Omega_{K,0}| \lesssim  10^{-3}$  \cite{Ade:2013ktc}. This also serves as an estimate of the contribution of more complex anomalies, which could generically be decomposed in a superposition of plane waves, suggesting that generalizing the underlying ansatz would be consistent with  the absence of any observed curvature.   

\begin{figure}[t]
\centering
\includegraphics[width=8cm,height=4.7cm]{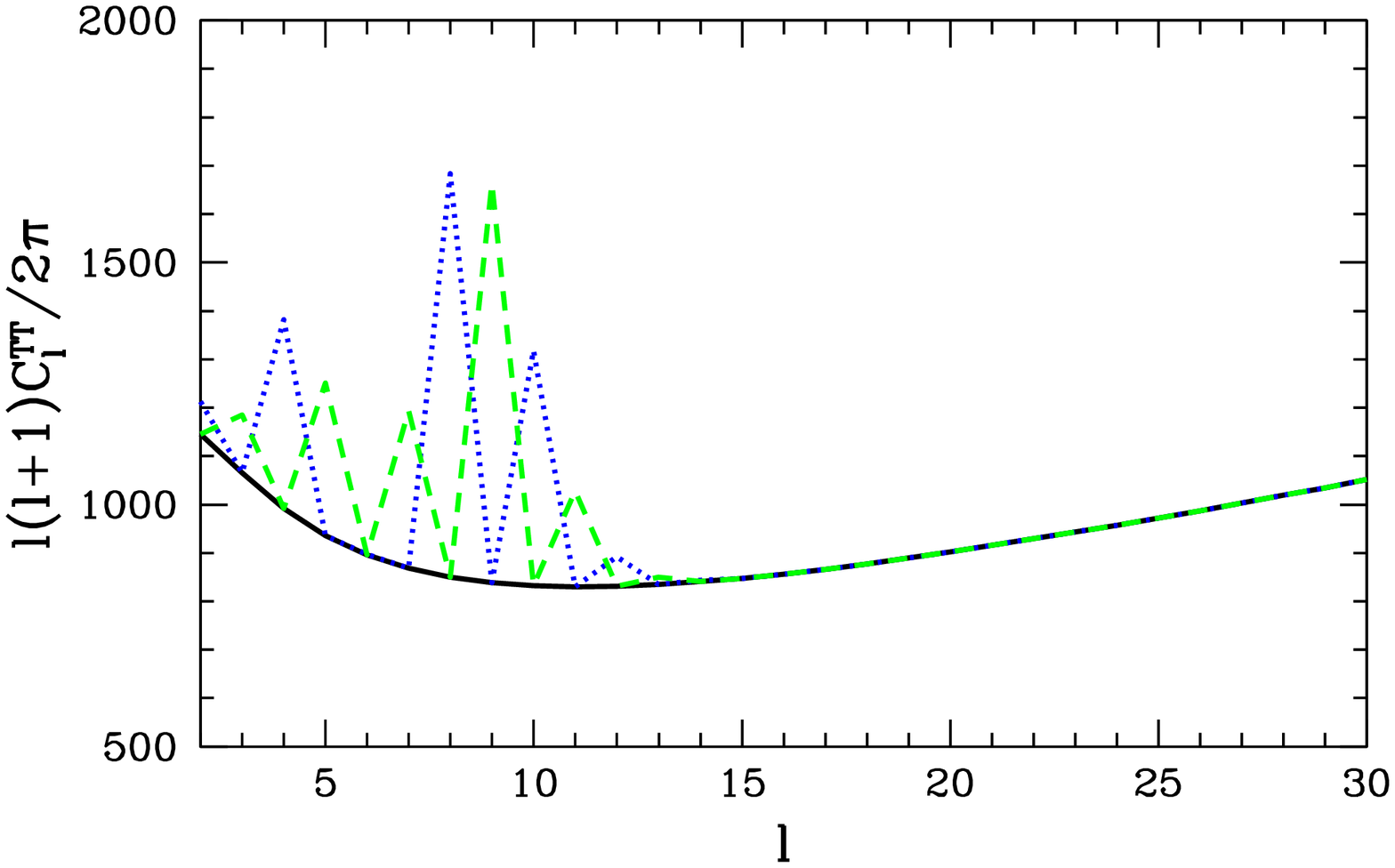}
\includegraphics[width=8cm,height=4.7cm]{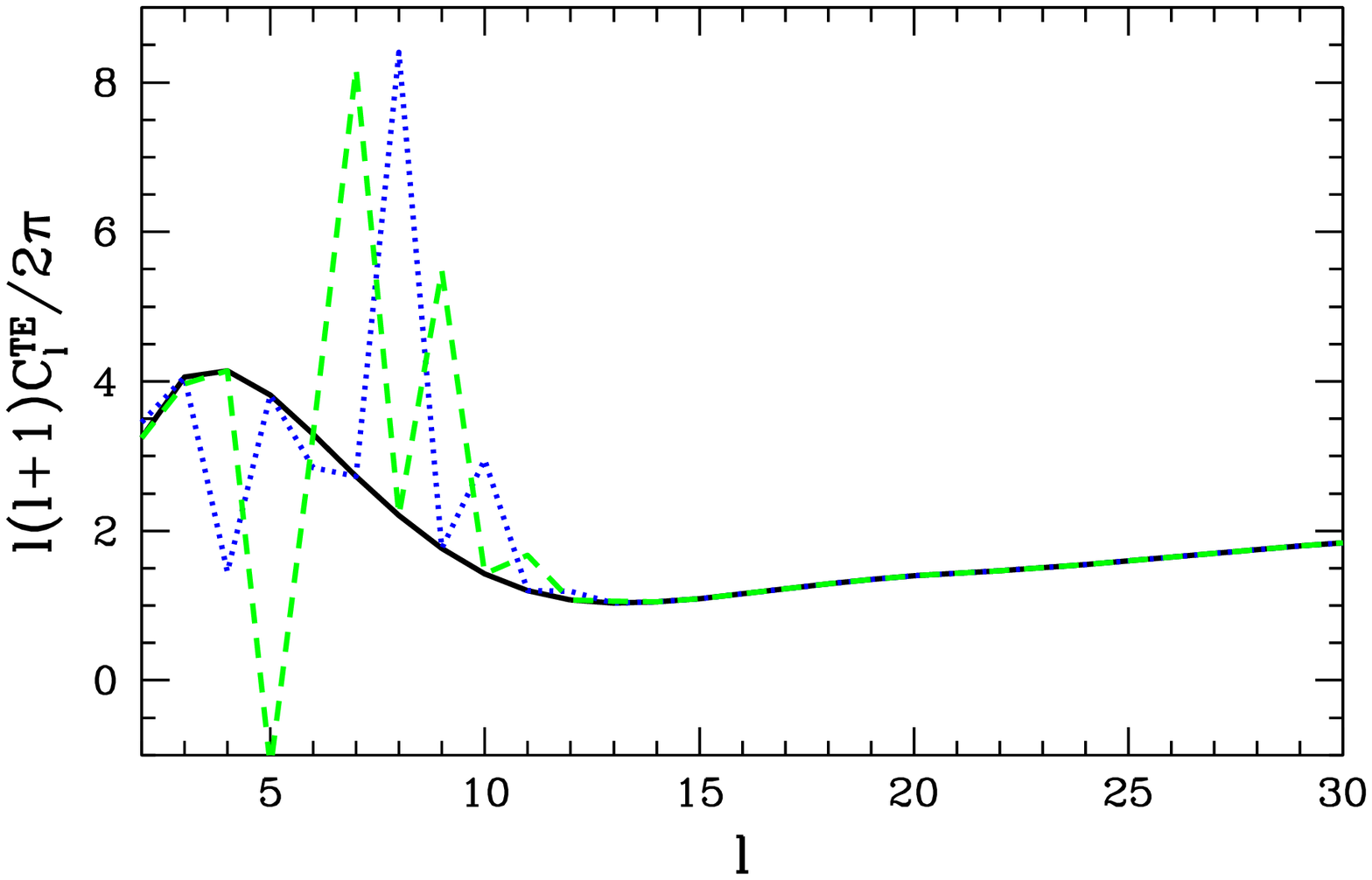}
 \includegraphics[width=8cm,height=4.7cm]{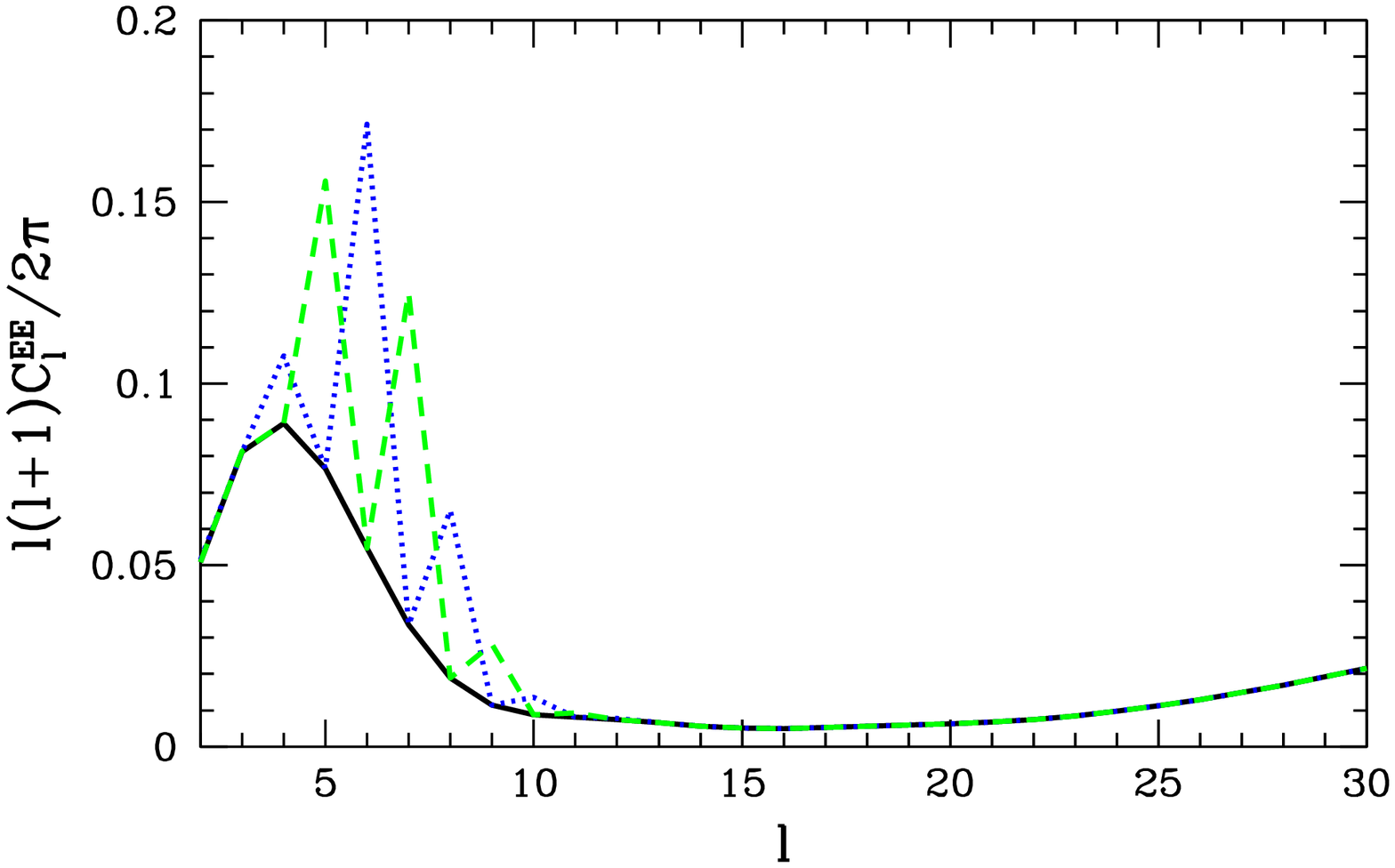}
\vspace{-10pt}
\caption{\label{cl_fig} $\langle TT \rangle$ (top), $\langle TE \rangle$ (middle),  and $\langle EE \rangle$ (bottom) power spectrum on large scales in $\mu\mathrm{K}^2$. The black solid curve corresponds to standard $\Lambda$CDM (no large scale perturbation), the blue dotted curve has a cosine perturbation added with  $A_f=3\times10^{-5}$, $\alpha=0$ (only even $l$ affected), the green dashed curve has a cosine perturbation added with the same amplitude but with $\alpha=\pi/2$ (only odd $l$ affected).
}
\vspace{-15pt}
\end{figure}
 
Decomposing the linear fluctuation into spherical harmonics we find \cite{Aslanyan:2013zs}
\begin{equation}\label{alm_semiclassical}
a_{lm}=2\pi A_f(-i)^lg_l(k_f)Y_{lm}^*(\mathbf{\hat{k}_f})[e^{i\alpha}+(-1)^le^{-i\alpha}]\,,
\end{equation}
where $g_l$ is the radiative transfer function for the corresponding mode ($T$ or $E$).  When   $\alpha=0$ the perturbation only appears on the even $l$ modes while for $\alpha=\pi/2$ only odd $l$ modes are nonzero.  Consequently, this scenario can lead to parity asymmetry in the CMB, a well-known candidate anomaly \cite{Ade:2013nlj,Land:2005jq,Kim:2010gf,Gruppuso:2010nd,Liu:2013wfa}.

Purported anomalies in the temperature maps sourced by atypical primordial density fluctuations will be accompanied by a correlated anomaly in $E$-mode polarization \cite{Dvorkin:2007jp,Mortonson:2009qv,Copi01102013}. Note that the polarization is not yet measured to the cosmic variance limit, so future polarization data will cross-check any physical model which was only constrained  with temperature data.

We calculate the $\langle TT\rangle$, $\langle TE\rangle$, and $\langle EE\rangle$ power spectra for our simple model, showing the results in Fig.~\ref{cl_fig}. The black solid curves show standard power spectra without any fluctuation, while the other two curves include a fluctuation  with  $A_f=3\times10^{-5}$ and wavelength $2\pi/k_f=0.6L_0$. The blue dotted curve has $\alpha=0$ and the green dashed curve has $\alpha=\pi/2$ (our notation assumes that $\mathbf{x}=0$ at our position).

\begin{figure}[t]
\centering
\includegraphics[width=4cm,height=2cm]{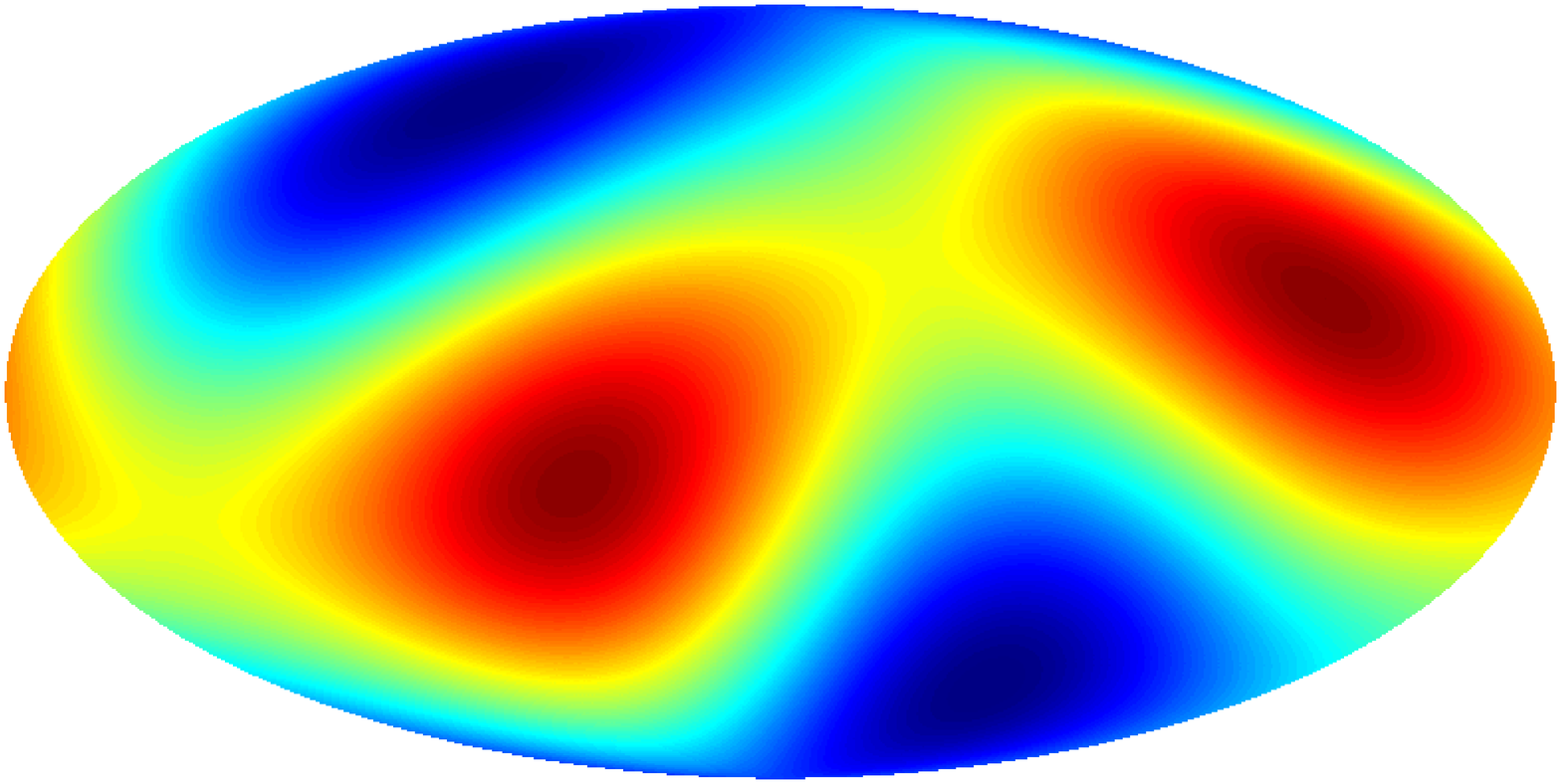}
\includegraphics[width=4cm,height=2cm]{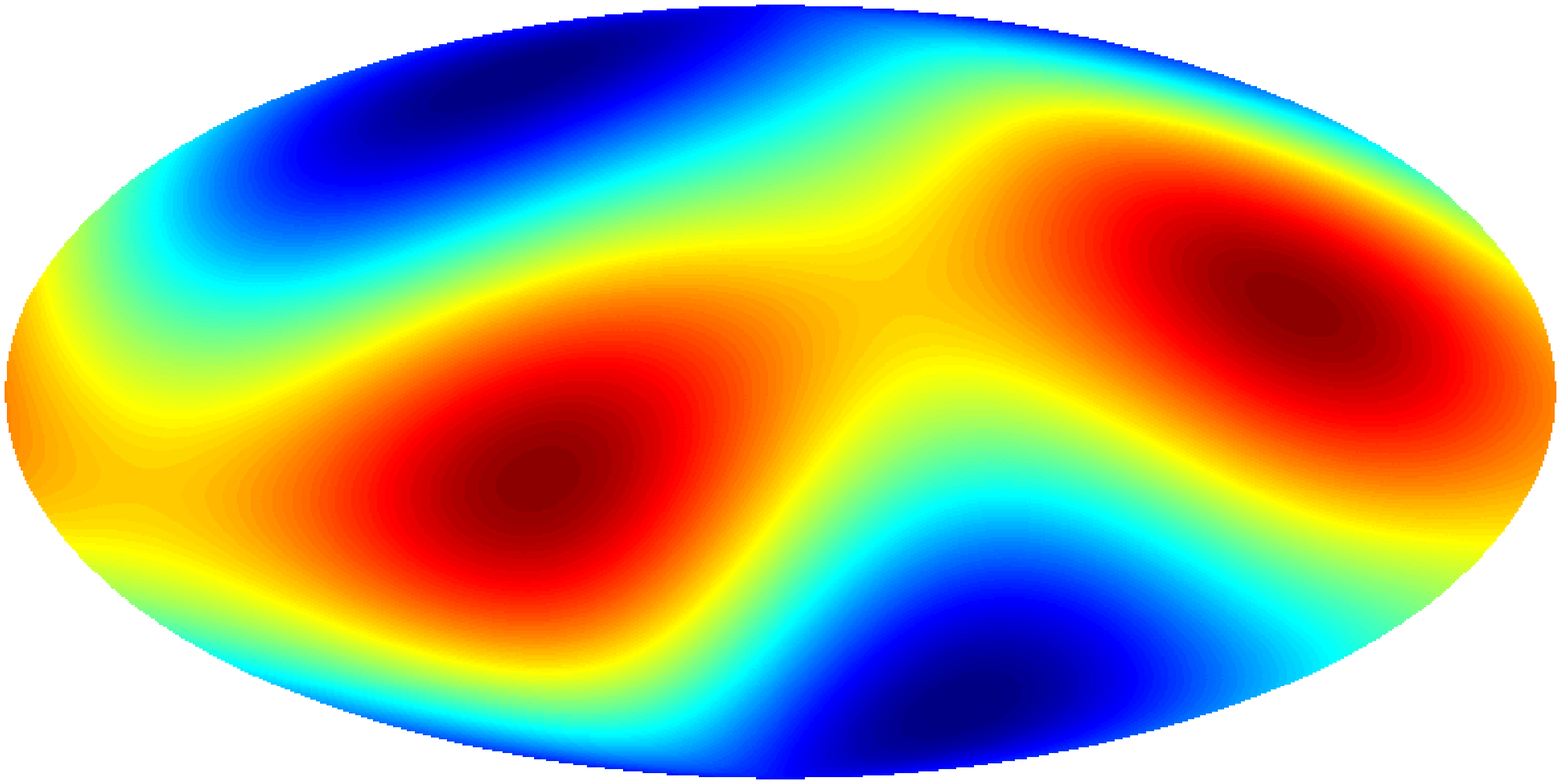}
\includegraphics[width=4cm,height=2cm]{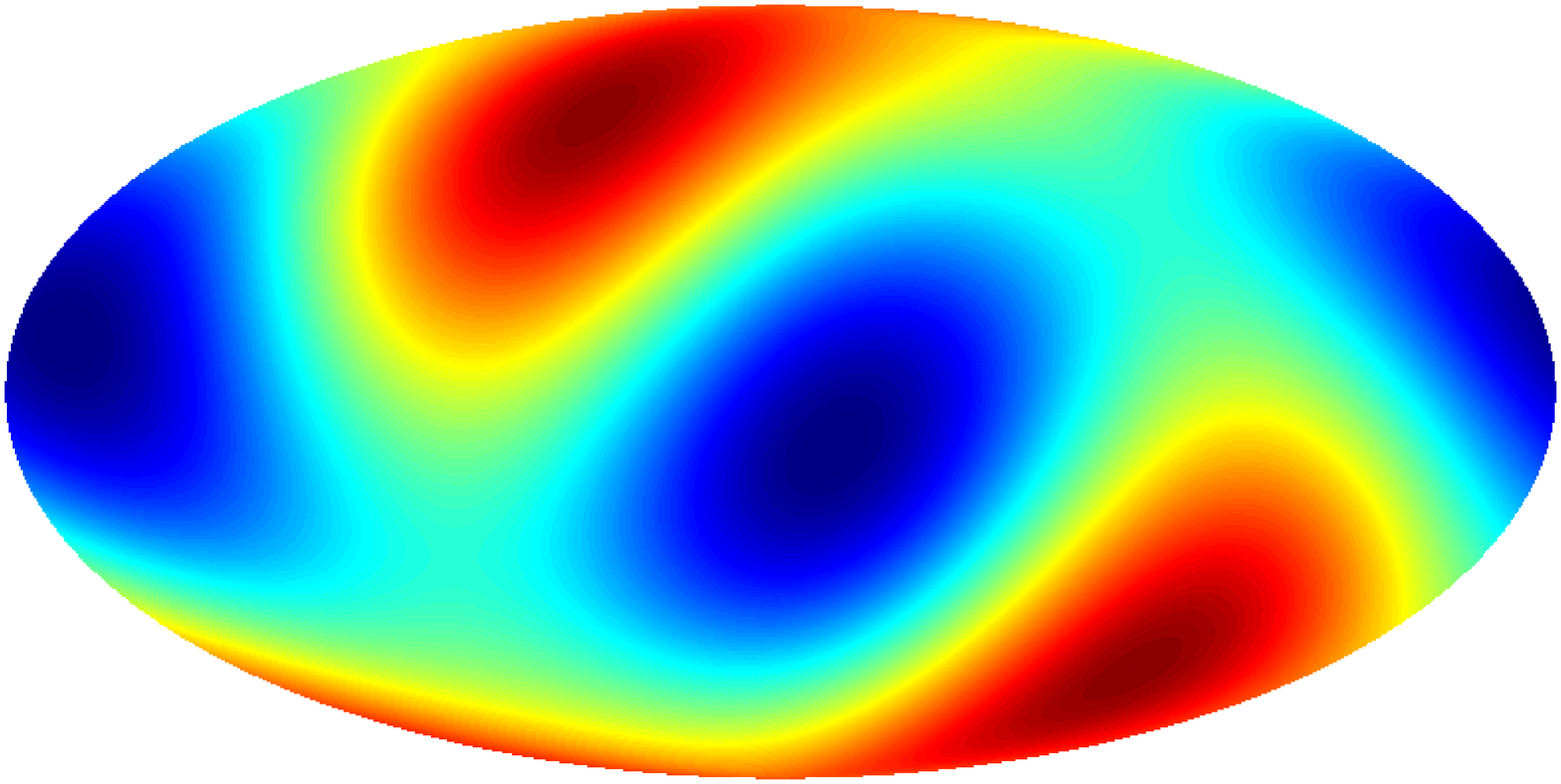}
\includegraphics[width=4cm,height=2cm]{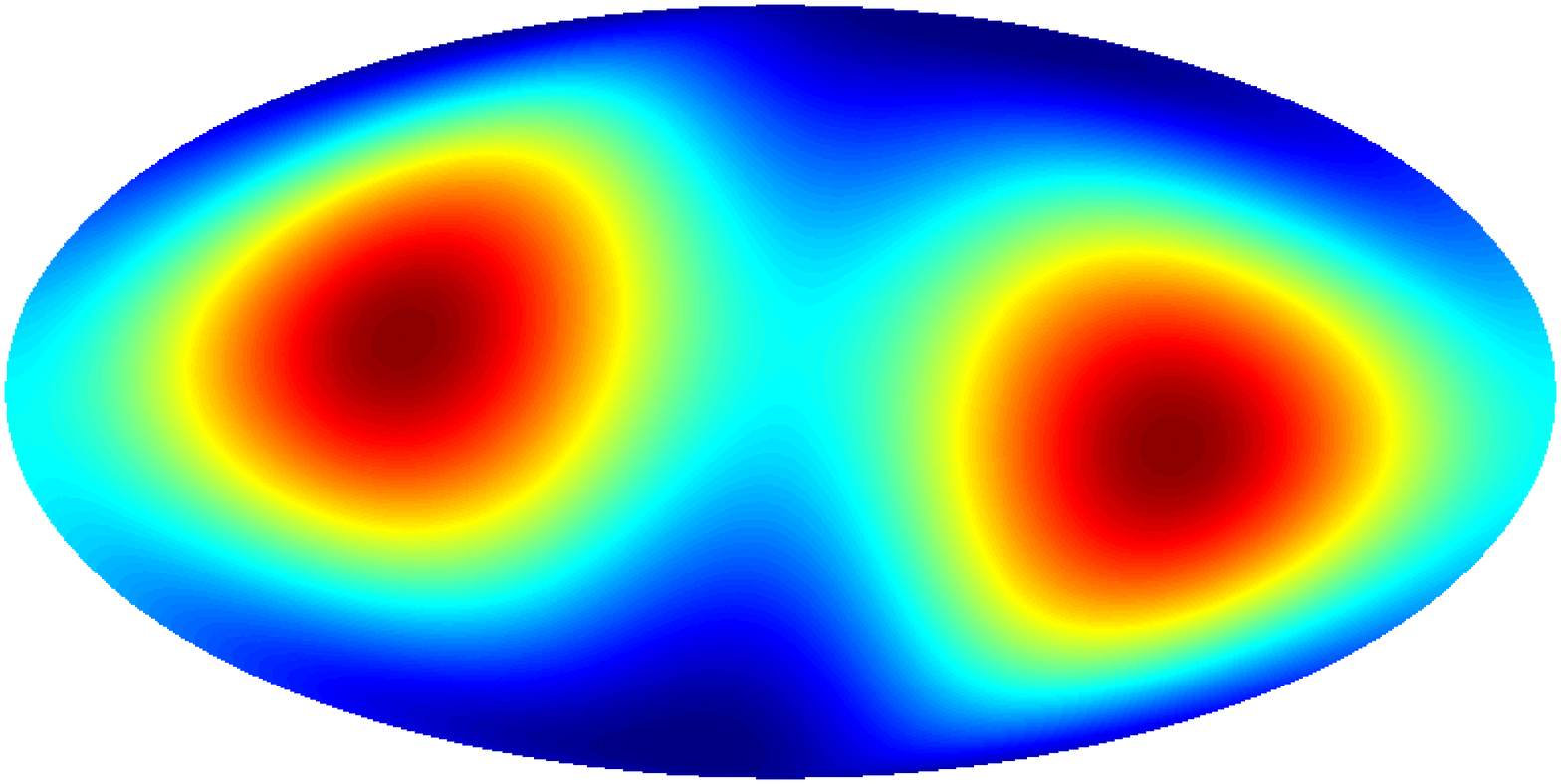}
\caption{\label{mode_alignment_fig} Angular momentum dispersion as a function of the axis direction for a simulation without (left) and with (right) an added cosine perturbation for $l=2$ (top) and $l=3$ (bottom).  The parameters used in this plot are $A_f=3\times10^{-5}$, $2\pi/k_f=0.6L_0$, $\alpha=\pi/4$.
}
\vspace{-12pt}
\end{figure}

Our candidate inhomogeneity can also align  low-$l$ modes in the CMB. The so-called ``axis of evil''  \cite{Ade:2013nlj,Land:2005ad,Tegmark:2003ve,deOliveiraCosta:2003pu,Rakic:2007ve} is usually analyzed by finding the axis $\mathbf{\hat{n}}$ around which the angular momentum dispersion $\sum_m m^2|a_{lm}(\mathbf{\hat{n}})|^2$ is maximized \cite{deOliveiraCosta:2003pu}, where $a_{lm}(\mathbf{\hat{n}})$ are the spherical harmonic coefficients in a coordinate system where the z axis has $\mathbf{\hat z} || \mathbf{\hat{n}}$. From Eq. (\ref{alm_semiclassical}) one can show that for a single linear fluctuation the angular momentum dispersion is maximized around axes perpendicular to the direction of the fluctuation for all $l$. Consequently, a single excited Fourier mode enhances the probability of multipole alignment. Moreover, for a fluctuation in two (nonparallel) Fourier modes the direction of maximum angular momentum dispersion will be given by the intersection of the corresponding planes. In Fig. \ref{mode_alignment_fig} we show the quadrupole and the octupole for a simulated standard sky both without a fluctuation and with a sinusoidal fluctuation in the $z$ direction and the addition of the fluctuation pushes the otherwise random directions of the multipoles toward the plane perpendicular to $z$.

\begin{figure}[t]
\centering
\includegraphics[width=8cm,height=4.7cm]{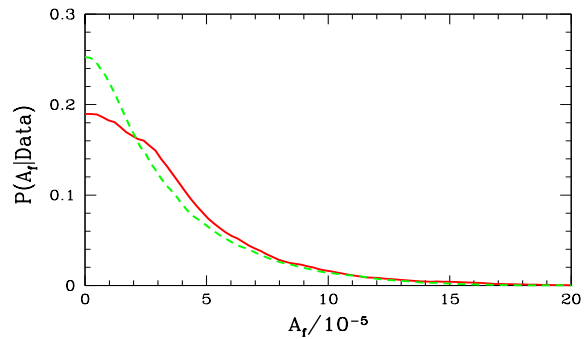}
\vspace{-10pt}
\caption{\label{amplitude_pdf_fig} The posterior probability distribution for $A_f$ from a simulated standard $\Lambda$CDM sky (no large scale perturbation added). The red solid curve is obtained from temperature data only, the green dashed curve includes polarization data as well.
}
\vspace{-13pt}
\end{figure}

The power spectra are proportional to $A_f^2$, so this signal is most efficiently constrained using the full likelihood function, which is sensitive to $A_f$. This analysis was performed in Ref.~\cite{Aslanyan:2013zs} using the WMAP temperature data.  To estimate the improvement expected from Planck polarization data, we repeat that analysis with a simulated standard universe (no fluctuation added), but including polarization data, with Planck noise \cite{:2006uk} in the likelihood function. The posterior probability distribution for the amplitude $A_f$ is shown in Fig. \ref{amplitude_pdf_fig}. The solid red curve is obtained from temperature data only, the green dashed curve includes polarization data. The $68.3\%$ upper bound on $A_f$ is $4.3\times10^{-5}$ from temperature data only, $3.7\times10^{-5}$ with polarization data included.

Considering a simulated sky that  includes a large scale fluctuation, we can assess the ability of polarization data to confirm the presence of the primordial inhomogeneity.  A fluctuation just below the threshold of detectability, with $A_f=3\times10^{-5}$ ($2\pi/k_f=L_0$, $\alpha=0$), yields an improvement in $\chi^2$ compared to the zero-amplitude case of $5.86$ from temperature data only, and $7.55$ with polarization data included.  With $A_f=5\times10^{-5}$ the $\chi^2$ changes by $17.64$ with temperature only and $25.81$ with polarization included.  Consequently, while polarization data does not significantly tighten a null result, it would be key to confirming an apparent detection. 

The linear perturbation discussed here was fit to the  WMAP7 data set in Ref.~\cite{Aslanyan:2013zs}. The best-fit and posterior probability distribution peaks near $\alpha=1.3$,  $\lambda_f=2\pi/k_f =1.0\,L_0$, and  $b=63^\circ$, $l=5^\circ$ for the direction $\mathbf{\hat{k}_f}$, and for this combination of $\alpha$ and $\lambda$ the power spectra have a strong odd-even asymmetry for small values of $l \lesssim 8$. As expected, the odd modes are enhanced compared to even modes since $\alpha$ is close to $\pi/2$ and matches the conjectured direction of the odd-even asymmetry \cite{Ade:2013nlj,Land:2005jq,Kim:2010gf,Gruppuso:2010nd}. However, the  best-fit direction is not close to being perpendicular to the axis of evil direction ($b=60^\circ$, $l=-100^\circ$) \cite{Ade:2013nlj,Land:2005ad,Tegmark:2003ve,deOliveiraCosta:2003pu,Rakic:2007ve}, so the best-fit fluctuation does not make a significant contribution to the mode alignment. However, for $\alpha$ such that only  odd or even modes are strongly enhanced only modes with a matching parity will be aligned. Consequently, to account for the observed mode alignment we would need more than one planar inhomogeneity.

Parameter estimation for a planar perturbation is computationally expensive \cite{Aslanyan:2013zs}, and Bayesian evidence  has not been computed for this scenario. However,  $\Delta \chi^2 \approx 12$ for the best fit, improving the maximum likelihood by $e^6\approx 400$ while adding 5 free parameters. The improved likelihood is restricted to a subregion of the overall parameter volume, suggesting that the marginalized likelihood is not  significantly larger than that of  $\Lambda$CDM  and  the odds ratio is not significantly different from unity \cite{Jeffries1998}. Consequently, the planar model is unlikely to be favored by Bayesian evidence, despite a  nontrivial $p$ value associated with large scale parity violation.  

Modifications to the primordial perturbations do not only affect the CMB, but also modify  density fluctuations  in the current epoch.  The amplitude of the perturbation described by Eq.~(\ref{my_model})  is of the same order of magnitude as the usual quantum fluctuations, leading to  signatures in large scale structure surveys that are potentially observable but which do not completely undermine global homogeneity.  For the best-fit  wavelength of $\lambda_f=1.0\,L_0$, the density fluctuation varies from $0$ to its peak amplitude over $L_0/4\approx3.6\,\text{Gpc}$, which maps a distance in redshift space of $z\sim1$. Consequently,  probes of large scale structure at $z\gtrsim1$ such as the Planck SZ cluster catalogue \cite{Ade:2013skr} could further constrain the model. 

Other observational consequences of a local density variation include the KSZ effect,  Compton $y$ distortion \cite{Bull:2013fga,Valkenburg:2012td}, and anisotropic cosmic expansion \cite{Kalus:2012zu}.  For our model, matter density fluctuations at  $z=0$ are of order $10^{-4}$, implying a similar variation in the spacetime metric and the expansion rate. This is  within the current observational limit \cite{Bull:2013fga,Valkenburg:2012td,Kalus:2012zu}, although the detailed bounds were derived for models that differ from ours, so this is not a detailed comparison.  However, it is clear that CMB anomalies generated by  large scale inhomogeneities will have correlated signatures in large scale structure data, adding further discriminatory power to the approach described here.

In this Letter we contrast the {\em a posteriori} analysis of large scale anomalies in the microwave background with treatments based on a specific, physical model of the early Universe.  In this case we have a causal model of the mechanism that underlies the anomaly and  can  examine correlations between  observables and  anomalies.  We derived an order of magnitude estimate for   the amplitude of $\Omega_K$ induced by a large fluctuation and showed that large scale ``anomalies'' in the CMB need not also induce apparent spatial curvature.  
We analyzed  correlations between temperature and polarization data for a simple model, and the ability of polarization data to confirm an apparent anomaly in the temperature maps.  This specific model can explain the parity asymmetry of CMB without violating the flatness of the Universe at a detectable level. 

 Our results suggest that apparent anomalies in the microwave background can be usefully understood via the large scale fluctuations that might generate them. Moreover, anomalies in the microwave sky itself are effectively selected {\em a posteriori\/} from an infinite and ill-defined set of possible anomalies. Consequently these analyses suffer from a large and unquantified look-elsewhere effect which is not accounted for by $p$ values and similar frequentist statistics. By contrast, physical scenarios can be assessed  in terms of their intrinsic credibility, described via systematic expansions, and tested using model selection tools.
 
 \vspace{.3cm}

We thank Eugene Lim, Aneesh Manohar, Hiranya Peiris, Layne Price, and Glenn Starkman for useful comments. The authors acknowledge the contribution of the NeSI high-performance computing facilities and the staff at the Centre for eResearch at the University of Auckland. New Zealand's national facilities are provided by the New Zealand eScience Infrastructure (NeSI) and funded jointly by NeSI's collaborator institutions and through the Ministry of Business, Innovation and Employment's Infrastructure programme \cite{nesi}.

\bibliography{citations}

\end{document}